\pdfoutput=1
\documentclass[english,fleqn,twoside]{article}
\usepackage[T1]{fontenc}
\usepackage[latin1]{inputenc}
\usepackage{color}
\usepackage{graphicx}

\providecommand{\tabularnewline}{\\}

\usepackage[headings]{espcrc2}

\usepackage{color}

\title{Heavy Quarks:  Lessons Learned from HERA and Tevatron\thanks{To 
appear in the proceedings of  the 
Ringberg Workshop, 
New Trends in HERA Physics 2008; 
October 5 --- 10, 2008, 
Ringberg Castle, Tegernsee.
}}

\author{Fredrick Olness,\address[SMU]{Southern Methodist University, Dallas, TX 75275-0175 USA}\thanks{Presented by Fred Olness}
Ingo Schienbein,\address[CRNS]{Laboratoire de Physicque Subatomique et de Cosmologie, Universit\'e
Joseph Fourier  Grenoble 1,
CRNS/IN2P3, Institut National Polytechnique de Grenoble, 38026 Grenoble,  France} 
}

\runtitle{Heavy Quarks}
\runauthor{Olness \& Schienbein}

\usepackage{babel}

\begin{document}
\begin{abstract}

We review some of the recent developments which have enabled the heavy quark mass to be incorporated into both the calculation of the 
hard-scattering cross section and the PDFs. We compare and contrast some of the schemes that have been used in recent global PDF analyses, and 
look at  issues that arise when these calculations are extended to NNLO.

\end{abstract}
\maketitle

\def\lsim{\mathrel{\hbox{\rlap{\hbox{\lower4pt\hbox{$\sim$}}}\hbox{$<$}}}} 
\def\gsim{\mathrel{\rlap{\lower4pt\hbox{\hskip1pt$\sim$}} \raise1pt\hbox{$>$}}} 

\tableofcontents{}\vspace{1cm}

\section{Introduction}

The production of heavy quarks in high energy processes has become
an increasingly important subject of study both theoretically and
experimentally. The theory of heavy quark production in perturbative
Quantum Chromodynamics (PQCD) is more challenging than that of light
parton (jet) production because of the new physics issues brought
about by the additional heavy quark mass scale. The correct theory
must properly take into account the changing role of the heavy quark
over the full kinematic range of the relevant process from the threshold
region (where the quark behaves like a typical {}``heavy particle'')
to the asymptotic region (where the same quark behaves effectively
like a parton, similar to the well known light quarks $\{u,d,s\}$).

We review theoretical methods which have been advanced to improve
existing QCD calculations of heavy quark production, and the impact
on recent experimental results from HERA and the Tevatron.

\begin{figure}[t]
\includegraphics[width=1\columnwidth]{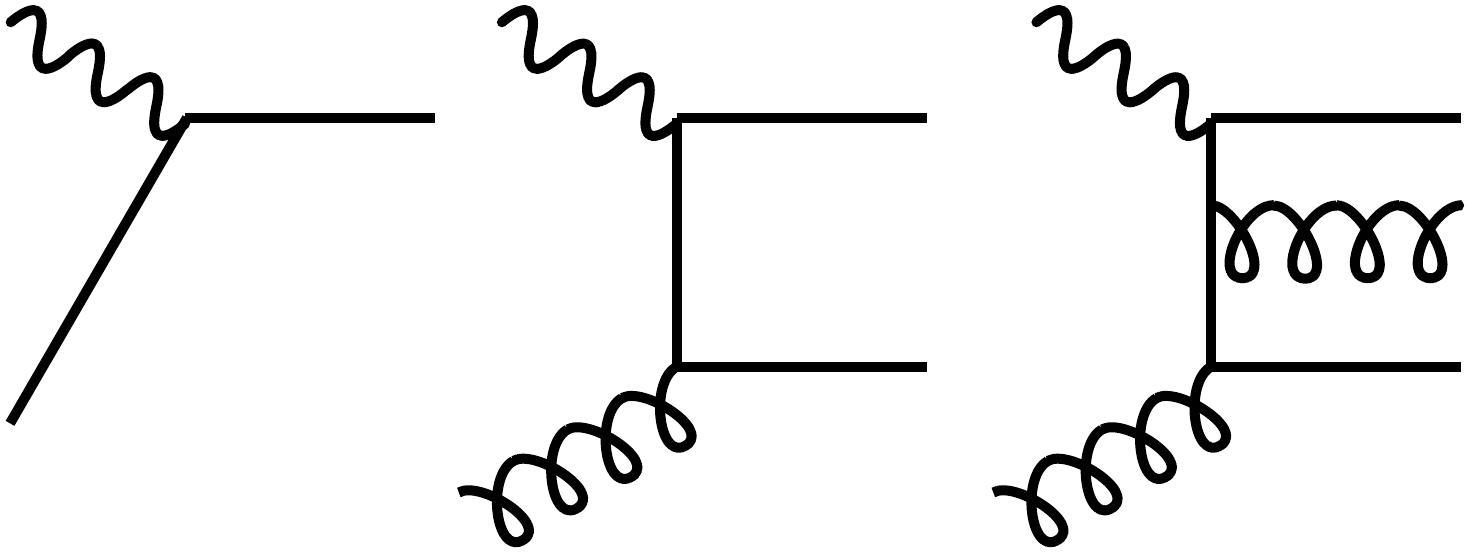}\vspace{-0.5cm}

\caption{Characteristic Feynman graphs which contribute to DIS heavy quark
production: a)~the LO ${\cal O}(\alpha_{S}^{0})$ quark-boson scattering
$QV\to Q$, b)~the NLO ${\cal O}(\alpha_{S}^{1})$ gluon-boson scattering
$gV\to Q\bar{Q}$, and c)~the NNLO ${\cal O}(\alpha_{S}^{2})$ boson-gluon
scattering $gV\to gQ\bar{Q}$. \label{fig:diagrams}}

\vspace{-0.5cm}

\end{figure}

The ACOT renormalization scheme provides a mechanism to incorporate
the heavy quark mass into the theoretical calculation of heavy quark
production both kinematically and dynamically. In 1998 Collins\cite{Collins:1998rz}
extended the factorization theorem to address the case of heavy quarks;
this work provided the theoretical foundation that allows us to reliably
compute heavy quark processes throughout the full kinematic realm.

\subsection{NLO DIS calculation}

Figure~\ref{fig:diagrams} displays characteristic Feynman graphics
for the first few orders of DIS heavy quark production. If we consider
the DIS production of heavy quarks at ${\cal O}(\alpha_{S}^{1})$
this involves the LO $QV\to Q$ process and the NLO $gV\to Q\bar{Q}$
process.

The key ingredient provided by the ACOT scheme is the subtraction
term (SUB) which removes the {}``double counting'' arising from
the regions of phase space where the LO and NLO contributions overlap.
Specifically, the subtraction term is: \[
\sigma_{SUB}=f_{g}\otimes\tilde{P}_{g\to Q}\otimes\sigma_{QV\to Q}\quad.\]
$\sigma_{SUB}$ represents a gluon emitted from a proton ($f_{g}$)
which undergoes a collinear splitting to a heavy quark $(\tilde{P}_{g\to Q})$
convoluted with the LO quark-boson scattering $\sigma_{QV\to Q}$.
Here, $\tilde{P}_{g\to Q}(x,\mu)=\frac{\alpha_{s}}{2\pi}\,\ln(\mu^{2}/m_{c}^{2})\, P_{g\to c}(x)$
where $P_{g\to c}(x)$ is the usual $\overline{MS}$ splitting kernel.

\subsection{When do we need Heavy Quark PDFs}

\begin{figure}[t]
\includegraphics[width=0.48\textwidth]{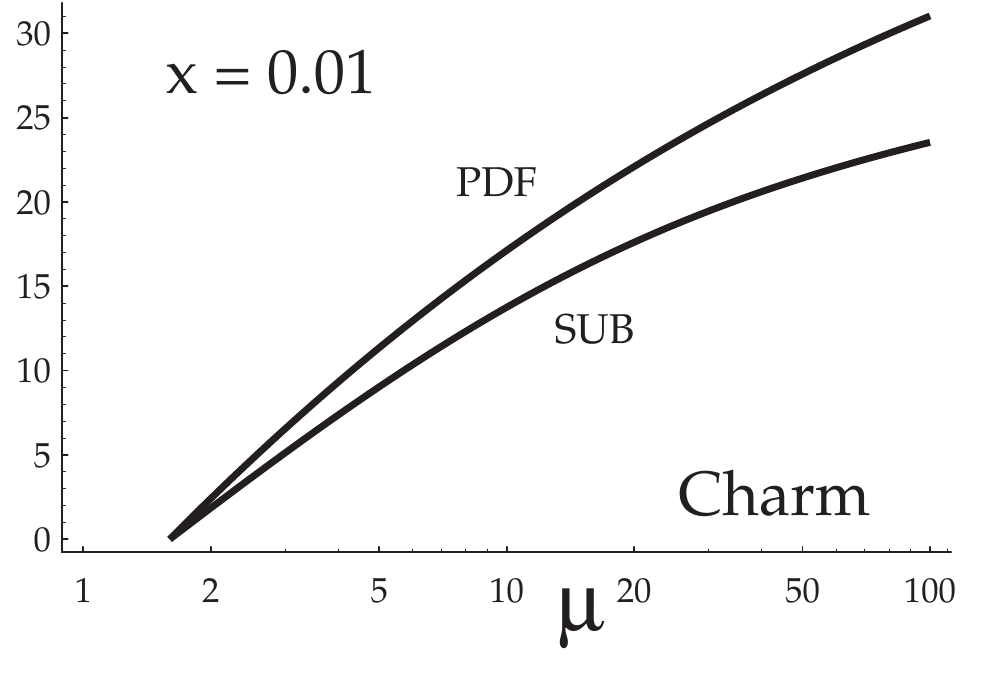}

\includegraphics[width=0.48\textwidth]{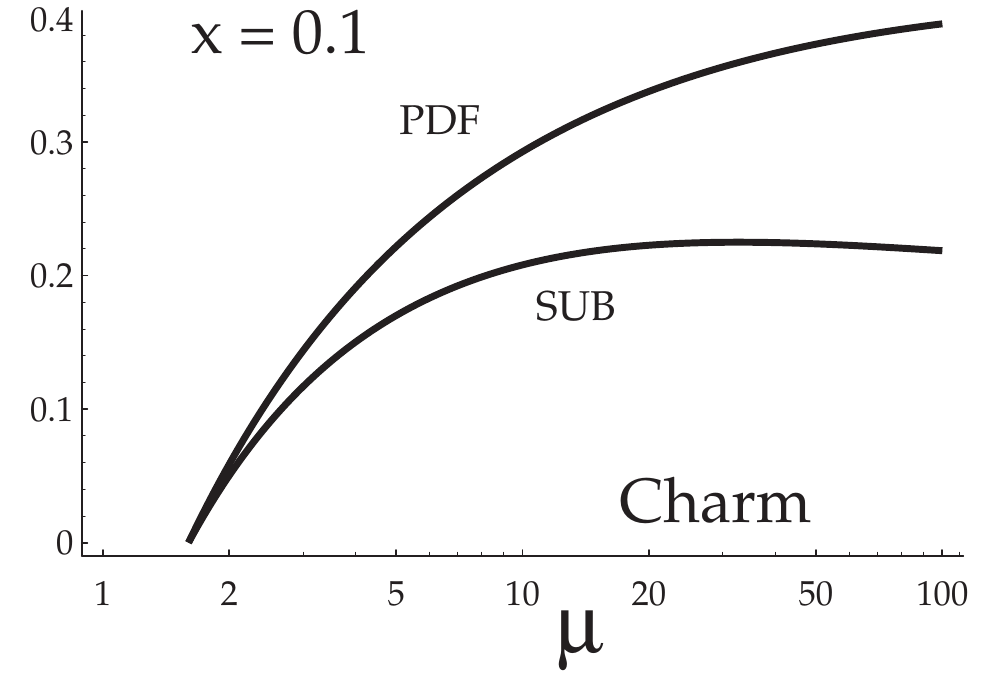}\vspace{-1cm}

\caption{Comparison of the DGLAP evolved charm PDF $f_{c}(x,\mu)$ with the
perturbatively computed single splitting (SUB) $\tilde{f}_{c}(x,\mu)=f_{g}(x,\mu)\otimes\tilde{P}_{g\to c}$
charm evolution vs. $\mu$ in GeV for two representative values of
$x$. \label{fig:charmPDF} }

\vspace{-0.5cm}

\end{figure}

The novel ingredient in the above calculation is the inclusion of
the heavy quark PDF contribution which resums logs of $\ln(\mu^{2}/m_{Q}^{2})$.
One can ask the question: When do we need to consider such terms?
The answer is illustrated in Figure~\ref{fig:charmPDF} where we
compare the DGLAP evolved PDF $f_{c}(x,\mu)$ with the single splitting
perturbative result

The DGLAP PDF evolution sums a non-perturbative infinite tower of
logs while the SUB contribution removes the perturbative single splitting
component which is already included in the NLO contribution. Hence,
at the PDF level the difference between the heavy quark DGLAP evolved
PDF $f_{Q}$ and the single-splitting perturbative $\tilde{f}_{Q}$
will indicate the contribution of the higher order logs which are
resummed into the heavy quark PDF. Here, we shall find it convenient
to define $\tilde{f}_{Q}=f_{g}\otimes\tilde{P}_{g\to Q}$ which represents
the PDF of a heavy quark $Q$ generated from a single perturbative
splitting. 

For $\mu\sim m_{Q}$ we see that $f_{Q}$ and $\tilde{f}_{Q}$ match
quite closely, whereas $f_{Q}$ and $\tilde{f}_{Q}$ differ significantly
for $\mu$ values a few times $m_{Q}$. While the details will depend
on the specific process, in general we find that for $\mu$ scales
3 to 5 times $m_{Q}$ the terms resummed by the heavy quark PDF can
be significant.

\section{\hbox{The ACOT Renormalization Scheme}}

\subsection{Massive vs. Massless Evolution}

\begin{figure*}[t]
\includegraphics[width=1\textwidth]{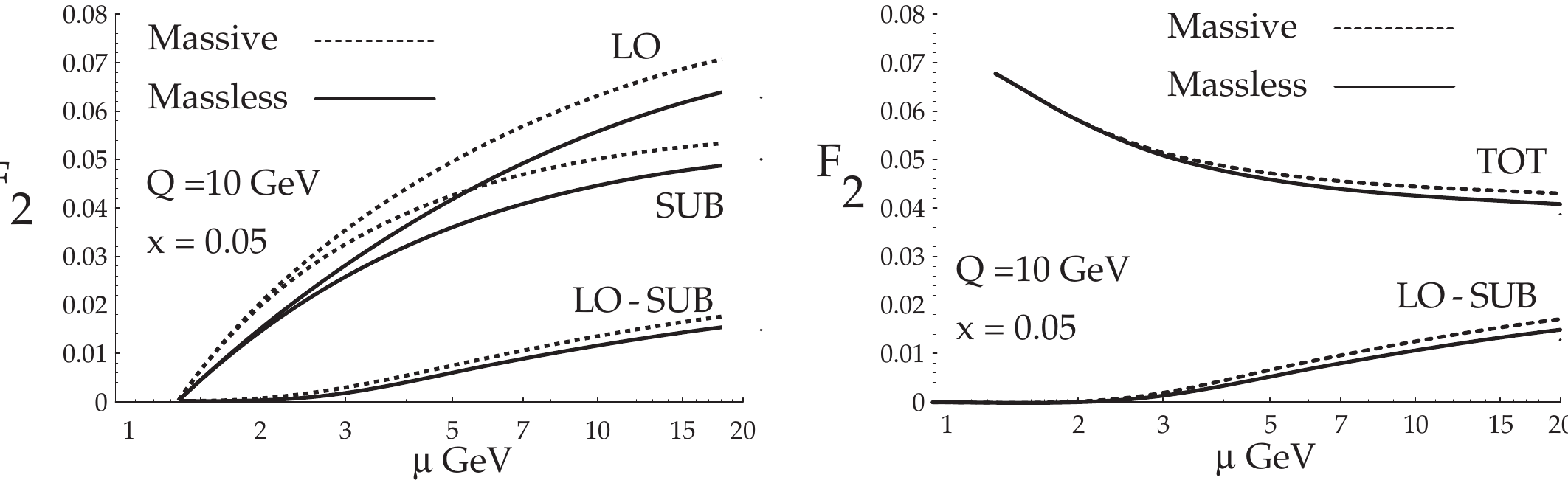}\vspace{-1.25cm}

\caption{Comparison of heavy quark DIS structure function for mass-dependent
(massive) and mass-independent (massless) evolution. \label{fig:massiveEvolution}}

\vspace{-0.25cm}

\end{figure*}

Another useful result that arises from the proof of Collins\cite{Collins:1998rz}
is that we can use mass-independent (massless) evolution kernels to
evolve the heavy quark PDFs without any loss of accuracy as compared
to a mass-dependent (massive) evolution kernel.\cite{Olness:1997yn}
Specifically, Collins demonstrated that consistent application of
the formalism correctly resums the massive contributions up to higher-twist
corrections ${\cal O}(\Lambda_{QCD}^{2}/Q^{2})$ and that there are
no errors of order ${\cal O}(m_{Q}^{2}/Q^{2})$.

This result is illustrated in Figure~\ref{fig:massiveEvolution}
where we compare the results of a NLO DIS heavy quark production calculation
using massless and massive DGLAP evolution kernels. In Fig.~\ref{fig:massiveEvolution}a)
we see that while the choice of massive or massless kernels significantly
changes the individual $LO$ and $SUB$ contributions, the difference
$LO-SUB$ which contributes to the total ($TOT=LO-SUB+NLO$) is minimal.
This numerically verifies that the choice of massive or massless evolution
kernels is purely a scheme choice which has no physical content.

While we see this result demonstrated numerically in Figure~\ref{fig:massiveEvolution},
the underlying reason for this result is closely related to the previous
observations made regarding Figure~\ref{fig:charmPDF}. The LO result
is given by $LO\sim f_{Q}\otimes\sigma_{Q\to Q}$ and the subtraction
term is given by $SUB\sim f_{g}\otimes\tilde{P}_{g\to Q}\otimes\sigma_{Q\to Q}$.
If we expand the DGLAP equation for $f_{Q}$ in the region $\mu\sim m_{Q}$
we find $f_{Q}\sim f_{g}\otimes\tilde{P}_{g\to Q}+{\cal O}(\alpha_{s}^{2})$;
thus, we have $LO\sim f_{g}\otimes\tilde{P}_{g\to Q}\otimes\sigma_{Q\to Q}+{\cal O}(\alpha_{S}^{2})$.
We observe that while $LO$ and $SUB$ individually depend on the
specific splitting kernels, the combination $LO-SUB$ is insensitive
to whether we use the massive or massless kernel.%
\footnote{While we have given a heuristic description of this result (in which
we used some illustrative approximations), we emphasize the proof
applies to all cases and does not require any such approximations. %
}

Therefore, we conclude that so long as the splitting kernels $P_{a\to b}$
are matched between the DGLAP evolution and the definition of the
subtractions (SUB), the choice of a massive or massless DGLAP evolution
kernel was purely a choice of scheme and the physical results are
invariant.

\subsection{S-ACOT}

In a complementary application, it was observed that the heavy quark
mass could be set to zero in certain pieces of the hard scattering
terms without any loss of accuracy. This modification of the ACOT
scheme goes by the name Simplified-ACOT (S-ACOT) and can be summarized
as follows. 
\begin{description}
\item [{S-ACOT:}] For hard-scattering processes with incoming heavy quarks
or with internal on-shell cuts on a heavy quark line, the heavy quark
mass can be set to zero ($m_{Q}=0$) for these pieces.\cite{Kramer:2000hn}
\end{description}
If we consider the case of NLO DIS heavy quark production, this means
we can set $m_{Q}=0$ for the LO terms ($QV\to Q$) as this involves
an incoming heavy quark, and we can set $m_{Q}=0$ for the SUB terms
as this has an on-shell cut on an internal heavy quark line. Hence,
the only contribution which requires calculation with $m_{Q}$ retained
is the NLO $gV\to Q\bar{Q}$ process.

\begin{figure}[t]
\includegraphics[width=1\columnwidth]{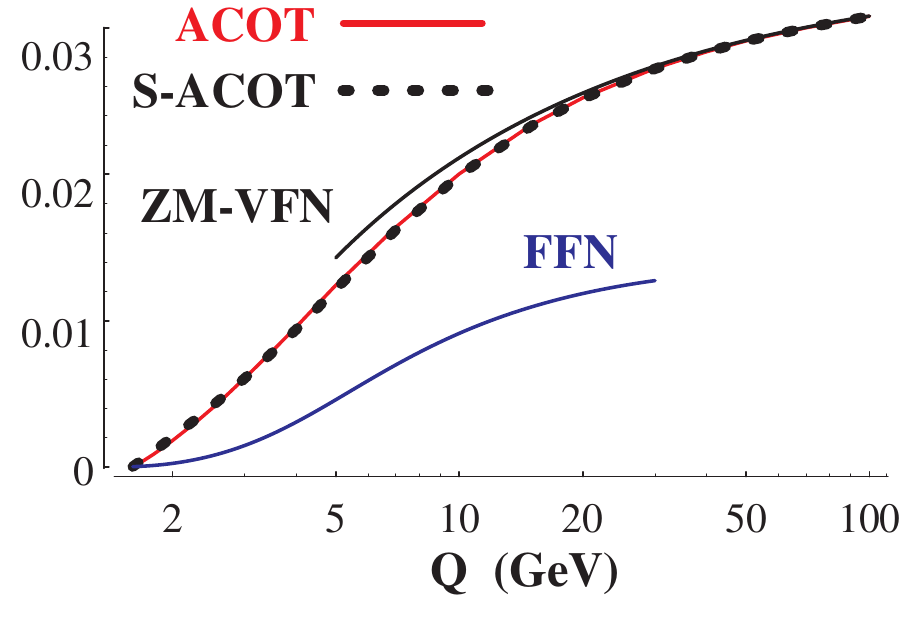}\vspace{-1cm}

\caption{Comparison of schemes for NLO DIS heavy quark production as a function
of $Q$. We display calculations using the ACOT, S-ACOT, Fixed-Flavor
Number (FFN), and Zero-Mass Variable Flavor Number (ZM-VFN) schemes.
The ACOT and S-ACOT results are virtually identical. \label{fig:sacot}}

\vspace{-0.5cm}

\end{figure}

Figure~\ref{fig:sacot} displays a comparison of a calculation using
the ACOT scheme with all masses retained vs. the S-ACOT scheme; as
promised, these two results match throughout the full kinematic region.

\subsection{ACOT-$\chi$}

In the conventional implementation of the heavy quark PDFs, we must
{}``rescale'' the Bjorken $x$ variable as we have a massive parton
in the final state. The original rescaling procedure is to make the
substitution $x\to x(1+m_{c}^{2}/Q^{2})$ which provides a kinematic
penalty for producing the heavy charm quark in the final state.\cite{Barnett:1976ak}
As the charm is pair-produced by the $g\to c\bar{c}$ process, there
are actually two charm quarks in the final state---one which is observed
in the semi-leptonic decay, and one which goes down the beam pipe
with the proton remnants. Thus, the appropriate rescaling is not $x\to x(1+m_{c}^{2}/Q^{2})$
but instead $x\to\chi=x(1+(2m_{c})^{2}/Q^{2})$; this rescaling is
implemented in the ACOT--$\chi$ scheme, for example.\cite{Amundson:1998zk,Amundson:2000vg,Tung:2001mv}
The factor $(1+(2m_{c})^{2}/Q^{2})$ represents a kinematic suppression
factor which will suppress the charm process relative to the lighter
quarks.

\subsection{Numerical Comparison}

\begin{figure}[t]
\includegraphics[width=0.85\columnwidth]{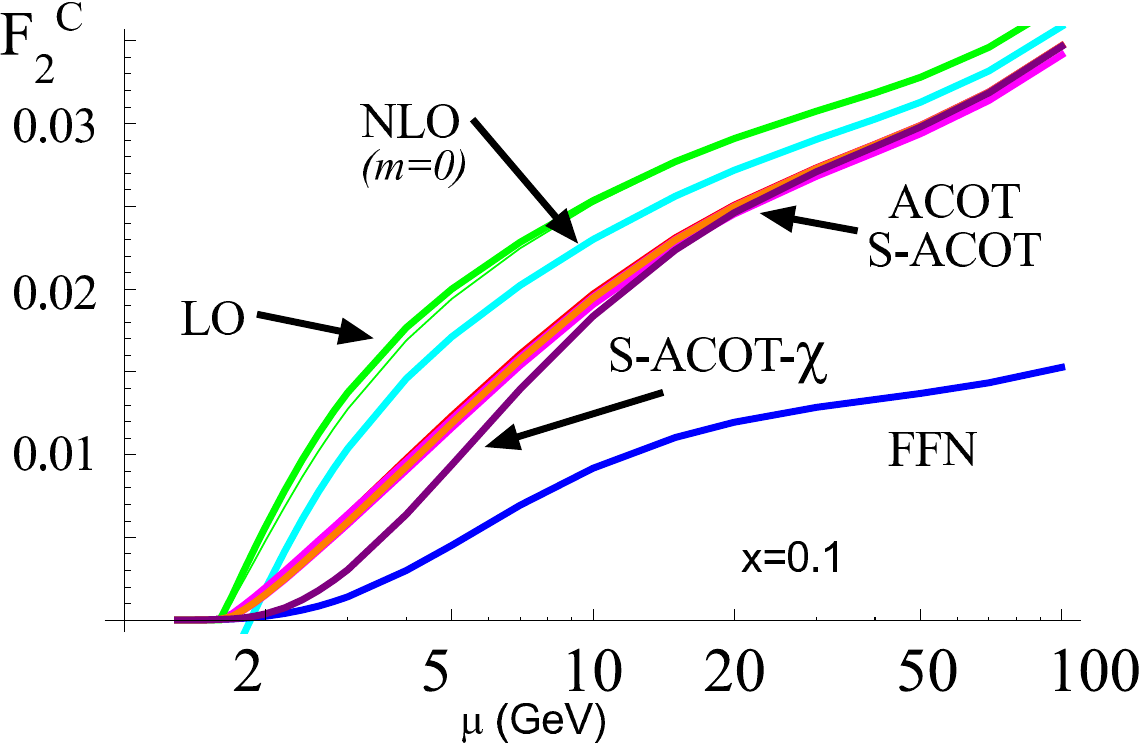}

\vspace{-1cm}

\caption{Calculation of DIS heavy quark production for a variety of schemes.
\label{fig:f2c}}

\vspace{-0.5cm}

\end{figure}

\begin{table*}[t]
\centering{}\begin{tabular}{|c|c||c|c||c|c|}
\hline 
\textbf{Set}  & \textbf{\# points}  & \textbf{CTEQ6HQ}  & \textbf{CTEQ6M}  & \textbf{\textcolor{red}{6M$\otimes$GM}}\textcolor{red}{ } & \textbf{\textcolor{red}{6HQ$\otimes$ZM}}\tabularnewline
\hline
\hline 
\textbf{ZEUS}  & 104  & 0.91  & 0.98  & \textcolor{red}{2.84 } & \textcolor{red}{3.72}\tabularnewline
\hline 
\textbf{H1}  & 484  & 1.02  & 1.04  & \textcolor{red}{1.50 } & \textcolor{red}{1.22}\tabularnewline
\hline
\hline 
\textbf{TOTAL}  & 1925  & 1.04  & 1.06  & \textcolor{red}{1.26 } & \textcolor{red}{1.30}\tabularnewline
\hline
\end{tabular}\caption{Table of $\chi^{2}$ per point for the individual HERA data sets,
and for the TOTAL of all data sets. (Non-HERA data sets are not displayed.)
The results are shown for CTEQ6HQ PDF using the General Mass (GM)
ACOT scheme, and CTEQ6M PDF using the zero-mass (ZM) $\overline{MS}$
scheme. We note the increased $\chi^{2}$ for mixed schemes using
CTEQ6M with the GM ACOT scheme, and the CTEQ6HQ with the ZM scheme.
\label{tab:chi2}}
\vspace{-0.3cm}

\end{table*}

Having introduced the various theoretical issues which enter the calculation
of the heavy quark process, we illustrate the numerical size of these
choices for the case of DIS heavy quark production.

In Figure~\ref{fig:f2c} we display the charm structure function
$F_{2}^{c}(x,\mu)$ for a variety of schemes and orders. $LO$ represents
the ${\cal O}(\alpha_{s}^{0})$ $QV\to Q$ process. $NLO$ includes
the ${\cal O}(\alpha_{s}^{1})$ processes (primarily $gV\to Q\bar{Q}$)
in the massless approximation. In the Fixed-Flavor-Scheme (FFS) the
heavy quark PDF is set to zero; hence, at ${\cal O}(\alpha_{s}^{1})$
this only receives contributions from $gV\to Q\bar{Q}$. The ACOT
and S-ACOT schemes are virtually identical---the curves are indistinguishable
in this plot. Finally, the implementation of the $\chi$-prescription
for the S-ACOT scheme (the ACOT-$\chi$ would yield identical results)
provides some additional suppression in the region $\mu\sim m_{Q}$.
To this order, our best theoretical estimate of the true cross section
would be either the ACOT-$\chi$ or equivalently S-ACOT-$\chi$.

To see the effect of these different results in the context of a global
fit we display the results for the CTEQ6M and CTEQ6HQ PDFs sets in
Table~\ref{tab:chi2}. Both the fits using a consistent application
of the ACOT and $\overline{MS}$ schemes yield good results. In contrast,
if we mismatch the scheme used in the PDF with that in the cross section
calculation we observe a dramatic increase in the $\chi^{2}$ values
obtained. This result underscores the importance of using properly
matched calculations.

\subsection{Heavy Quarks at the Tevatron }

\begin{figure}[t]
\includegraphics[width=1\columnwidth]{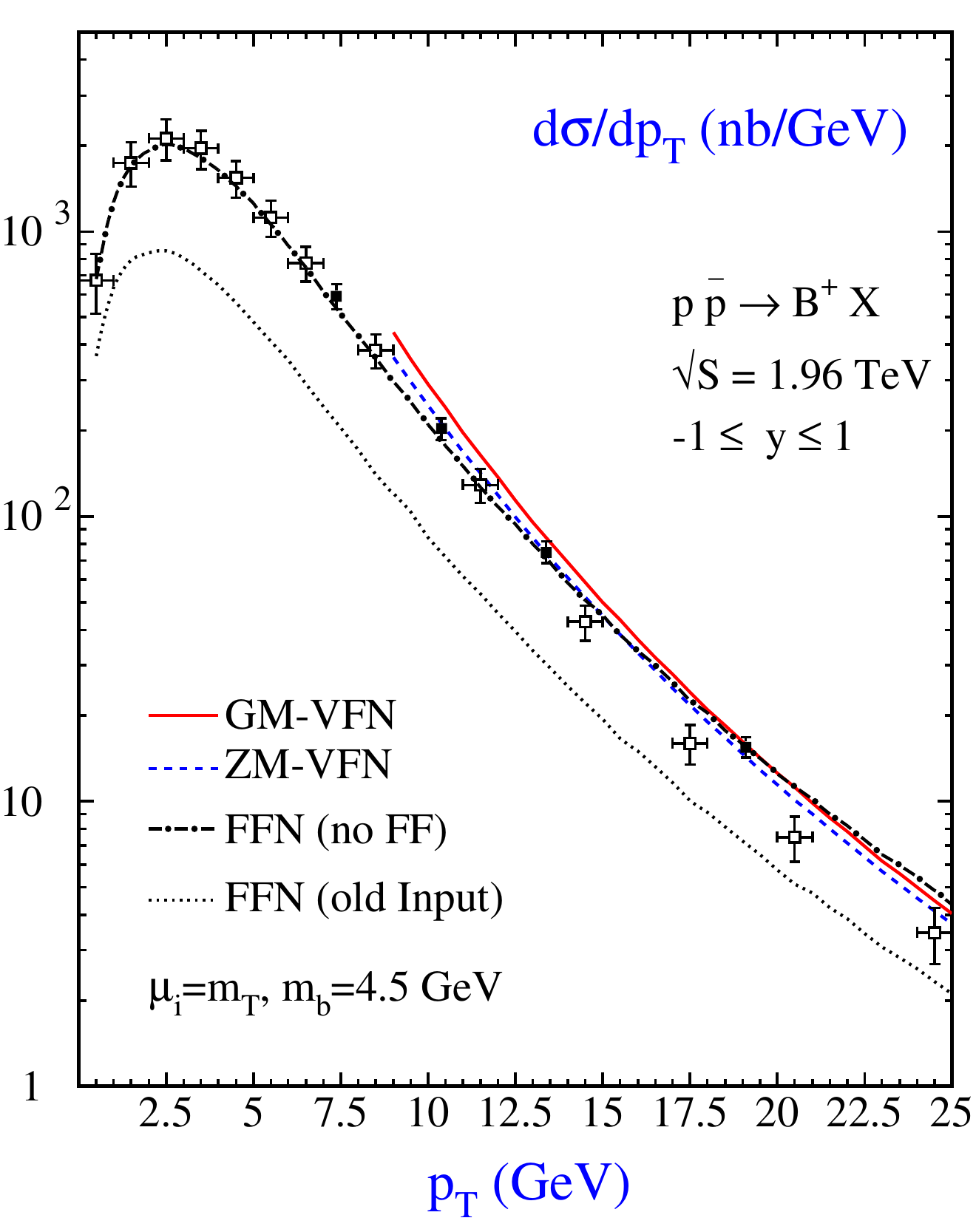}\vspace{-1cm}

\caption{From Ref.~\cite{Kniehl:2008zza}, the transverse momentum distribution
$d\sigma/dp_{T}$ for $p\bar{p}\to BX$ at $\sqrt{s}$=1.96\,TeV.
The results are shown for the General Mass (GM) Variable Flavor Number
(VFN) scheme and the Zero Mass (ZM) Variable Flavor Number (VFN) scheme.
Additionally, results are shown for the Fixed Flavor Number (FFN)
scheme with both recent PDFs (dot-dashed line) and the historical
PDFs (dotted line). The data is from the CDF collaboration.\cite{Abulencia:2006ps,Acosta:2004yw}
\label{fig:bprod}}

\vspace{-0.5cm}

\end{figure}

In the previous discussion we have primarily focused on DIS production
of heavy quarks for illustrative purposes as the formalism is easier
to layout when there is only a single hadron in the initial state.
Nevertheless, the same principles that we have used in the the DIS
case can be applied to that of the hadron-hadron initial state as
appropriate for the Tevatron and the LHC.

Historically, the predictions of b-production at hadron-hadron colliders
have been a challenge; the early results from the Tevatron were a
factor of 2 to 3 larger than the theoretical predictions. NLO QCD
corrections to the LO $gg\to Q\bar{Q}$ process were formidable and
yielded large corrections.\cite{Nason:1987xz,Altarelli:1988qr,Beenakker:1988bq,Nason:1989zy,Beenakker:1990maa}
It is interesting to observe that if the heavy quark PDF is taken
into account so that the LO contribution consists of both $gg\to Q\bar{Q}$
and $gQ\to gQ$, then the computed NLO contributions (with appropriate
subtractions) are thereby reduced suggesting improved convergence
of the perturbation theory.\cite{Cacciari:1993mq,Olness:1997yc,Cacciari:1998it,Cacciari:2001td,Cacciari:2002pa,Cacciari:2003uh}

Ref.~\cite{Kniehl:2008zza} performs a systematic comparison of the
GM-VFNS and ZM-VFNS using results of an updated analysis of hadronic
b-production at the Tevatron. Figure~\ref{fig:bprod} displays these
results for the Tevatron in the central rapidity region as compared
with the CDF data.\cite{Abulencia:2006ps,Acosta:2004yw} The result
is that the finite mass effects moderately enhance the $p_{T}$ distribution
in the region $p_{T}\sim2m_{H}$ by about 20\%, and this enhancement
decreases at larger $p_{T}$. For intermediate to large $p_{T}$ values
($p_{T}>m_{H}$) the three calculations (GM-VFNS, ZM-VFNS, FFN) match
quite closely, and are in good agreement with the data. Conversely,
if we use the FFN result with the historic values for the PDF and
$\alpha_{S}$ we find this prediction is roughly a factor of 3 below
the data. 

The excellent agreement between data and theory for this process is
an important achievement and represents the culmination of many years
of effort by both the theoretical and experimental community.

\section{Schemes used for Global Analysis}

\begin{figure*}[t]
\begin{tabular}{|c|c|}
\hline 
\includegraphics[width=0.48\textwidth]{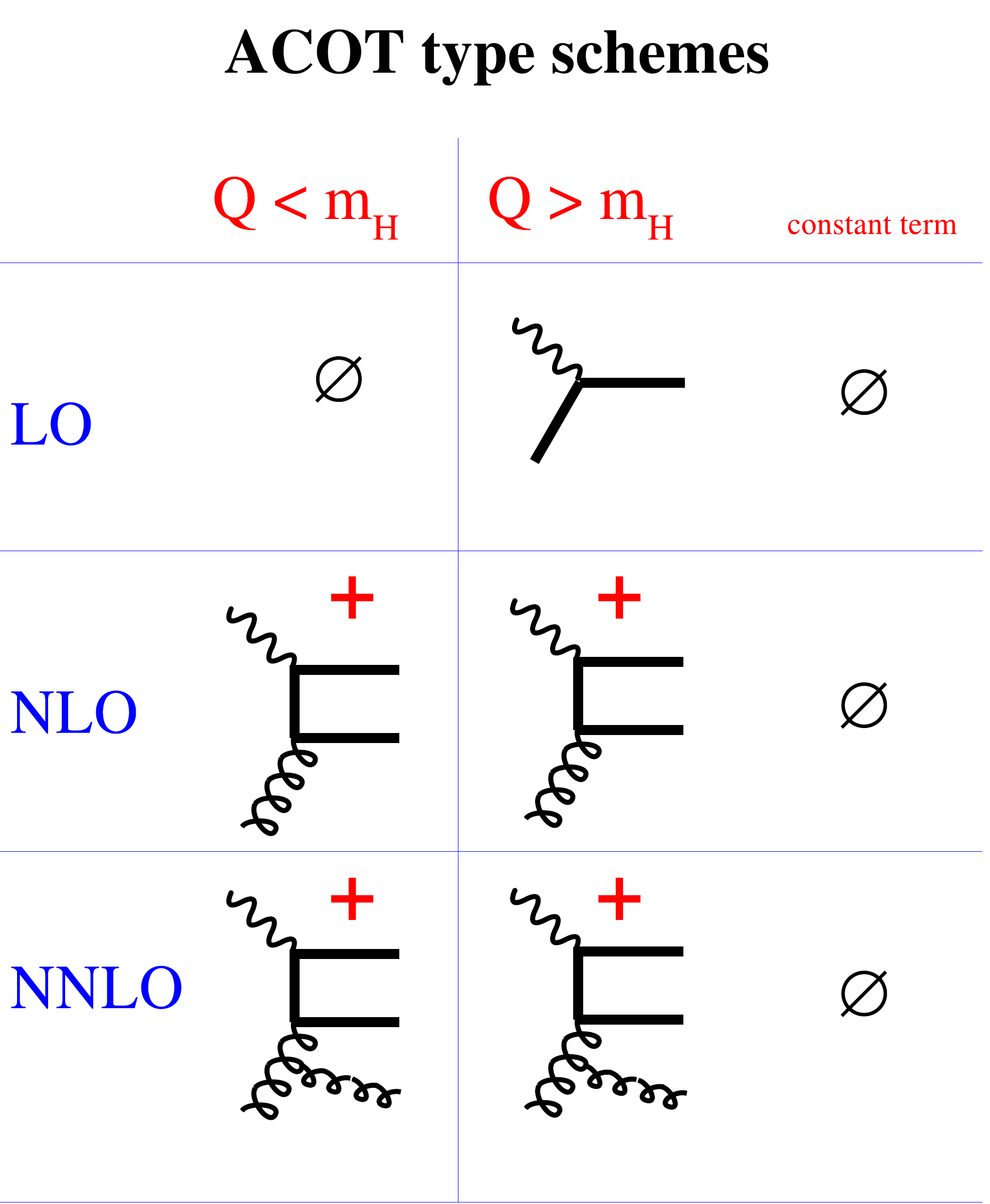}  & \includegraphics[width=0.48\textwidth]{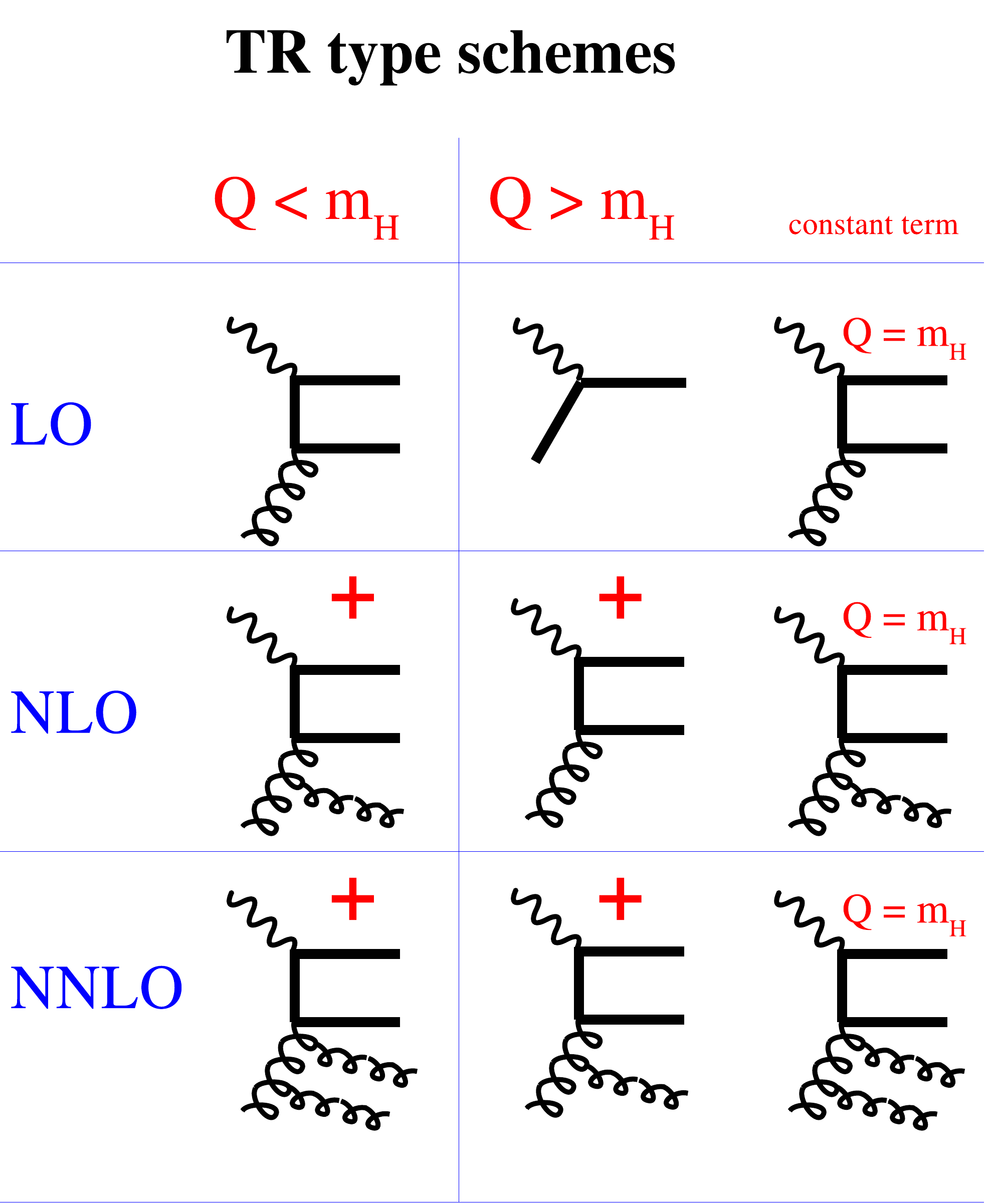}\tabularnewline
\hline
\end{tabular}\vspace{-0.75cm}

\caption{Diagrammatic comparison of TR and ACOT type schemes for the case of
DIS. This diagram is schematic to emphasize the similarities and differences.
The leading-order (LO) process is a ${\cal O}(\alpha_{s}^{0})$ boson
scattering from a heavy quark, e.g. $\gamma Q\to Q$; the NLO ${\cal O}(\alpha_{S}^{1})$
correction arises from $\gamma g\to Q\bar{Q}$, and the NNLO ${\cal O}(\alpha_{S}^{2})$
correction arises from $\gamma g\to Q\bar{Q}g$. \label{fig:schemes}}

\vspace{-0.25cm}

\end{figure*}

The ACOT scheme and variants were used for the CTEQ series of global
PDF fits.%
\footnote{Specifically, ACOT was used for CTEQ6HQ, and S-ACOT-$\chi$ was used
in CTEQ6.5 and CTEQ6.6. %
} For the MRST/MSTW series of global PDF fits the Thorne-Roberts (TR)
scheme was used. As these two sets of PDFs are widely used it is of
interest to compare and contrast these approaches. Figure~\ref{fig:schemes}
displays a diagrammatic comparison of the TR\cite{Thorne:2006qt,thorne:ringberg08}
and ACOT type schemes. While these schemes may appear quite different
at first glance, they differ by higher-order terms which will be reduced
as we increase the order of our perturbation theory.

In perturbation theory, we compute our observables to a fixed order
$N$ in $\alpha_{S}$; hence, we truncate the perturbation expansion
at ${\cal O}(\alpha_{S}^{N})$, and we have neglected terms of order
${\cal O}(\alpha_{S}^{N+1})$. In brief, the difference between these
two approaches amounts to adding different ${\cal O}(\alpha_{S}^{N+1})$
higher order terms. Thus, these two approaches will agree on the contributions
up to ${\cal O}(\alpha_{S}^{N})$. We will now review the motivation
and consequences of adding the differing higher order terms.

\subsection{Leading-Order (LO) $(\alpha_{S}^{0})$}

If we work at Leading-Order%
\footnote{Here, we define the order of the calculation according to the power
of $\alpha_{S}$; thus LO is $\alpha_{s}^{0}$, NLO is $\alpha_{s}^{1}$,
etc. %
} (LO) $\alpha_{S}^{0}$, when the heavy quark PDF is an {}``active''
parton (typically $\mu>m_{H})$ the LO contribution is $\gamma+Q\to Q$.
However, when the heavy quark PDF is \textbf{\emph{not}} an {}``active''
parton (typically $\mu<m_{H})$ the LO contribution vanishes. For
the ACOT scheme, no higher order terms are added to this results.
Hence for scales $\mu<m_{H}$, the LO answer is zero and we expect
large corrections to this result at NLO. For the TR scheme, a portion
of the $\gamma g\to Q\bar{Q}$ contribution is added; for $\mu<m_{H}$
the full $\gamma g\to Q\bar{Q}$ term is included, and for $\mu>m_{H}$
the $\gamma g\to Q\bar{Q}$ term frozen at $\mu=Q$ to avoid any difficulty
with large logarithms of the form $\ln(m_{H}/\mu)$.

Consequently, in the $\mu<m_{H}$ region the TR scheme yields a finite
LO result while the ACOT scheme yields zero. While both schemes formally
agree at ${\cal O}(\alpha_{S}^{0})$, clearly the ${\cal O}(\alpha_{S}^{1})$
terms can be important, particularly in the $\mu<m_{H}$ region.

\subsection{Next-to-Leading-Order (NLO) $(\alpha_{S}^{1})$}

If we work at NLO $(\alpha_{S}^{1})$, for the low $\mu$ region we
now include $\gamma g\to Q\bar{Q}$ as well as the $\gamma+Q\to Q$
process.%
\footnote{Note, in Figure~\ref{fig:schemes} and in the discussion the diagrams
an processes are schematic and illustrative. For example, at NLO we
include both $\gamma Q\to Qg$ and $\gamma g\to Q\bar{Q}$ as well
as all the corresponding subtractions. For details see Refs.~\cite{Thorne:2006qt,Aivazis:1993pi}.%
} If we again look in the region $\mu<m_{H}$, we find that while the
ACOT scheme yielded zero at LO, it now obtains a finite result at
NLO. For the TR scheme, in addition to the above terms, a portion
of the $\gamma g\to gQ\bar{Q}$ contribution is added; again, for
$\mu>m_{H}$ the $\gamma g\to gQ\bar{Q}$ term is frozen at $\mu=Q$
to avoid any difficulty with large logarithms.

As before, both the TR scheme and ACOT scheme formally agree at ${\cal O}(\alpha_{S}^{1})$,
but they will differ by the separate NNLO ${\cal O}(\alpha_{S}^{2})$
terms that have been included. In contrast to the LO case where the
ACOT scheme yielded zero for $\mu<m_{H}$, both schemes give finite
results in all kinematic region; hence, the relative difference will
be reduced.

\subsection{General Comparisons at Order $\alpha_{S}^{N}$}

Let us make some observations regarding these schemes at a general
order in perturbation ${\cal O}(\alpha_{S}^{N})$. We observe that
for a given set of processes calculated to $\alpha_{s}^{N}$, we can
implement the TR scheme to ${\cal O}(\alpha_{S}^{N-1})$ and the ACOT
scheme to ${\cal O}(\alpha_{S}^{N})$. For example, at NLO we note
that the ACOT scheme involves only graphs of order $\alpha_{s}^{1}$
while TR utilizes graphs of order $\alpha_{s}^{2}$. At present we
know the ${\cal O}(\alpha_{S}^{2})$ massive neutral current process
($\gamma g\to gQ\bar{Q}$, $\gamma Q\to ggQ$ and associated graphs);
hence, this allows us to compute the TR scheme to ${\cal O}(\alpha_{S}^{1})$
and the ACOT scheme to ${\cal O}(\alpha_{S}^{2})$. In contrast, the
massive charged current process is known only to ${\cal O}(\alpha_{S}^{1})$;
hence, this allows us to compute the TR scheme to ${\cal O}(\alpha_{S}^{0})$
and the ACOT scheme to ${\cal O}(\alpha_{S}^{2})$.

We note that recent improvements of theoretical techniques have enabled
significant advances in the calculation higher-order heavy quark processes.
For example, Ref.~\cite{Blumlein:2006mh} has obtained the asymptotic
results for $F_{L}^{Q\bar{Q}}(x,\mu)$ at the 3-loop order, and recently
Ref.~\cite{Bierenbaum:2008dk} has extended this work for the case
of $F_{2}^{Q\bar{Q}}(x,\mu)$. 

In general, the TR scheme achieves in practice the same highest asymptotic
order as ACOT by some modeling of terms below $Q^{2}=m_{Q}^{2}$ which
become (relatively) unimportant at high $Q^{2}$. As we move to higher
order calculations, the differences between these schemes will be
reduced as they arise from uncalculated higher-order contributions.

\section{NNLO and Beyond}

\begin{figure}
\includegraphics[angle=-90,width=1\columnwidth]{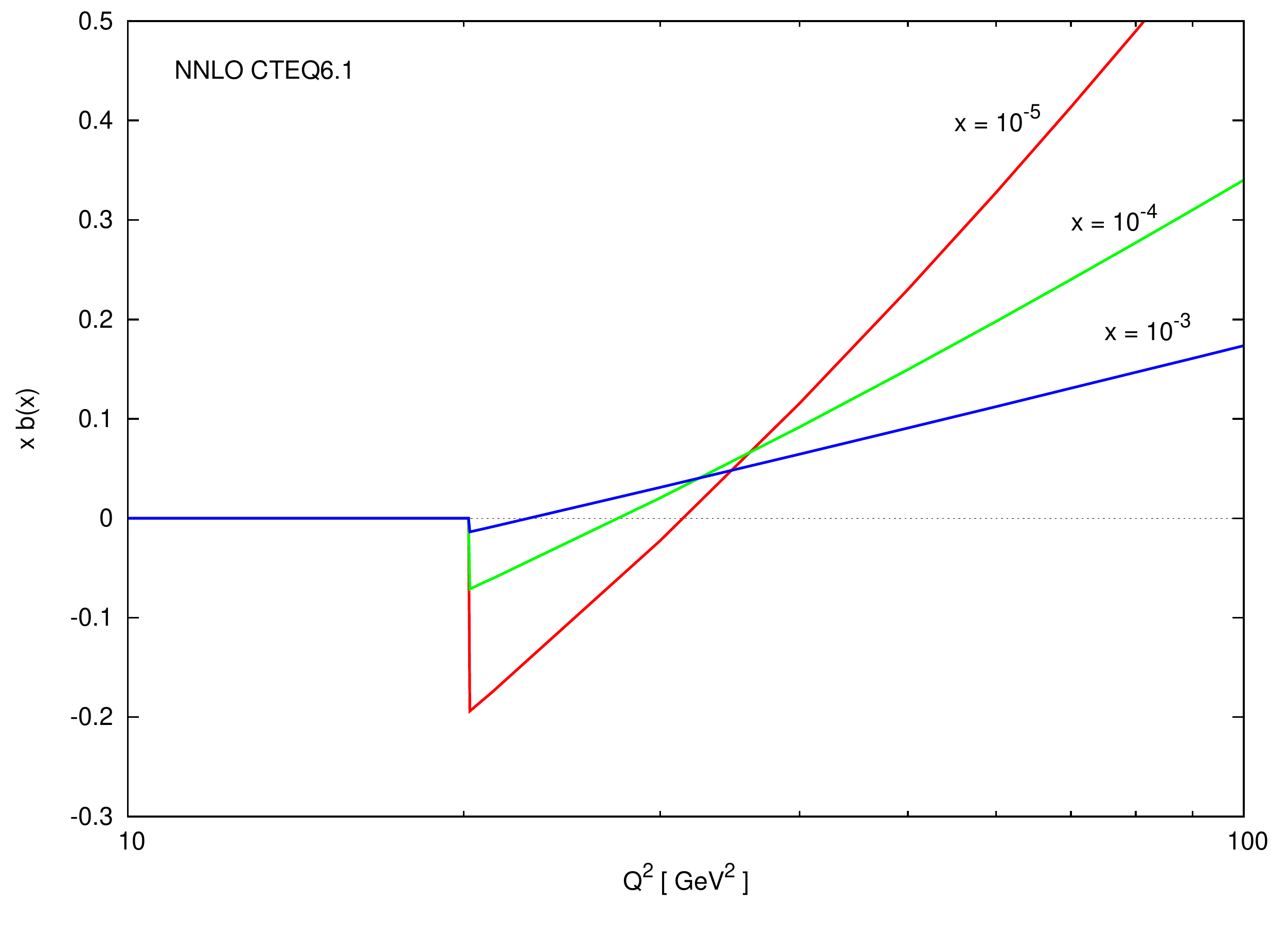}

\vspace{-1cm}
\caption{The b-quark PDF $x\, f_{b}(x,Q)$ with NNLO matching conditions for
3 choices of $x$. \label{fig:bpdf}}

\vspace{-0.75cm}

\end{figure}

\begin{figure}[t]
\includegraphics[width=1\columnwidth]{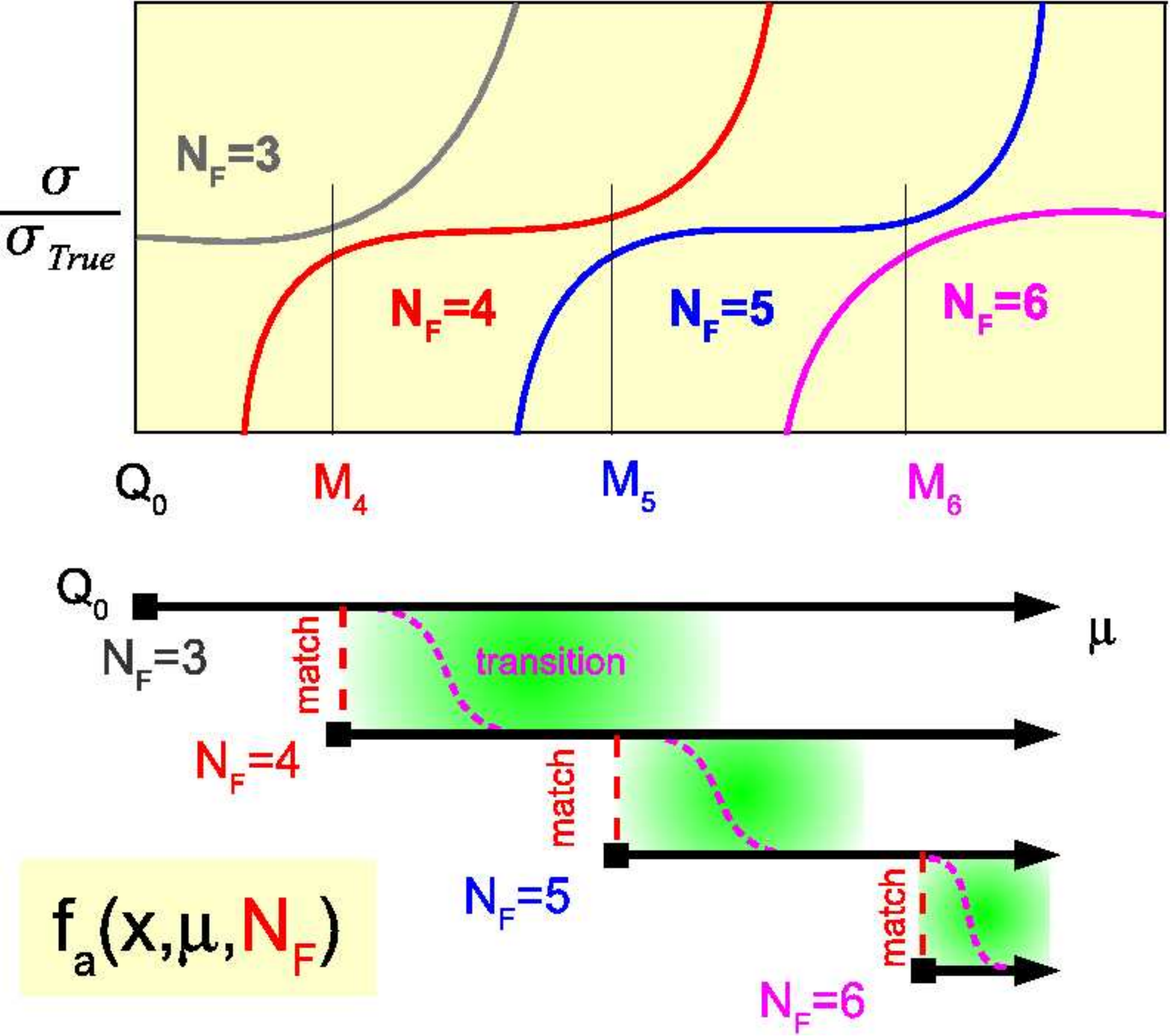}

\vspace{-1cm}

\caption{The upper figure schematically represents how each calculation with
a set number of flavors $N_{F}$ has a region of applicability. The
transition from the $N_{F}$-1 scheme to the $N_{F}$ scheme should
be in the vicinity of the $m_{N_{F}}$ mass, but need not occur exactly
at $\mu=m_{N_{F}}$. The lower figure illustrates that multiple PDFs
can co-exist for $\mu\geq m_{N_{F}}$ with matching performed at $\mu=m_{N_{F}}$.
\label{fig:NNLO}}

\vspace{-0.75cm}

\end{figure}

Although NLO is the state-of-the-art for many calculations, improved
experimental precision demands that we strive toward a NNLO accuracy.
When we consider PDFs for heavy quarks at NNLO, there are a number
of new elements that enter. 

One consequence is that the PDFs are no longer continuous across the
heavy flavor threshold. Even more, when matching charm and bottom
across their thresholds, they start from negative values as illustrated
in Figure~\ref{fig:bpdf}. The matching conditions have been computed
by a number of groups,\cite{Buza:1995ie,Cacciari:2005ry} and at NNLO
PDFs will have discontinuities of order ${\cal O}(\alpha_{S}^{2})$
when we transition from $N_{F}$ to $N_{F}+1$ flavors. While we may
be uncomfortable with discontinuities in our PDFs, we are reminded
that the PDFs are not physical observables, but instead are only theoretical
constructs which depend on (arbitrary) renormalization schemes and
scales.%
\footnote{Recall $\alpha_{S}(\mu)$ is also an unphysical theoretical construct;
this has discontinuities across flavor-thresholds at order $\alpha_{s}^{3}$. %
}

At NLO, the point $\mu=m_{Q}$ is special because $f_{a}^{N_{F}}(x,m_{Q})=f_{a}^{N_{F}+1}(x,m_{Q})$;
this is because the constant term in the matching equation happens
to be zero at NLO. Because of this {}``accident'' it was common
to use $\mu=m_{Q}$ as both the Matching Point and the Transition
Point. 

At NNLO the point $\mu=m_{Q}$ no longer has these special properties
as the transition from $N_{F}$ to $N_{F}+1$ will necessarily have
discontinuities at any value of $\mu$; hence, it may be desirable
to choose the Matching Point and the Transition Point at different
values of $\mu$. As these two point are not usually distinguished,
let us highlight their key features. 
\begin{description}
\item [{Matching~Point~$\mu_{M}$:}] The value of $\mu$ where the $N_{F}+1$
scheme is defined in terms of the $N_{F}$ scheme by a relation of
the form: $f_{a}^{N_{F}+1}(x,\mu)=A_{ab}\otimes f_{b}^{N_{F}}(x,\mu)$.
\item [{Transition~Point~$\mu_{T}$:}] The value of $\mu$ where the
user chooses to transition from the $N_{F}$ scheme to the $N_{F}+1$
scheme. 
\end{description}
Figure~\ref{fig:NNLO} schematically represents how each calculation
with a set number of flavors $N_{F}$ has a particular region of applicability
where it is best suited to describe the {}``true'' physics. The
complete description of the physics throughout the full kinematic
range will therefore consist of a patchwork of schemes which are {}``sewn
together.'' 

\vspace{0.5cm}

\textbf{The Transition Point: }It is easy to imagine situations where
we would not want to automatically transition between schemes at $\mu=m_{Q}$.
For example, consider we are analyzing data in the range $\mu\in[2,5]$\,GeV.
The bulk of the range is in the $N_{F}=4$ flavor region as $\mu>m_{c}\sim1.3$\,GeV,
but a small portion of the range extends above the $N_{F}=5$ flavor
region as $\mu>m_{b}\sim4.5$\,GeV. In the region $\mu\in[4.5,5]$\,GeV
it would be inconvenient to be forced to transition to a $N_{F}=5$
scheme because 1) the b-quark clearly plays no substantive role in
this kinematic range, and 2) both the PDFs and $\alpha_{s}(\mu)$
will have discontinuities at $\mu=m_{b}$. 

Clearly it is more reasonable to have the option to work consistently
in a $N_{F}=4$ flavor scheme even for $\mu\gsim m_{b}$. If PDFs
were generated such that the $N_{F}=4$ and $N_{F}=5$ schemes co-exist
in the region $\mu\sim m_{b}$, then the user could select $N_{F}$
by choice. 

The lower portion of Figure~\ref{fig:NNLO} illustrates how this
might be implemented. The PDFs can be generated such that the $N_{F}$
scheme is available for all $\mu\geq m_{N_{F}}$. Thus, for $\mu=5$\,GeV
the user would have access to schemes with $N_{F}=\{3,4,5\}$ and
can select the scheme by specifying $N_{F}$ in addition to $\{x,\mu\}$.
Therefore, the user could analyze their $\mu\in[2,5]$\,GeV data
set consistently in a single $N_{F}=4$ scheme, and choose to transition
to the $N_{F}=5$ scheme at a higher $\mu$ value to be specified
by the user. 

\vspace{0.5cm}

\textbf{The Matching Point:} Although the Matching Point can be set
to any $\mu$ value in the region of $m_{N_{F}}$, we shall argue
that the choice $\mu_{M}=m_{N_{F}}$ is optimal. 

First, we note that the Matching Point should be at or below the Transition
Point ($\mu_{T}\geq\mu_{M}$) if we desire to avoid downward DGLAP
evolution (which can be unstable). Therefore, if we perform the matching
at the heavy quark mass we have the reasonable constraint: $\mu_{T}\geq\mu_{M}=m_{N_{F}}$. 

Second, the matching conditions which define $f_{a}^{N_{F}+1}$ in
terms of $f_{a}^{N_{F}}$ are of the form $f_{a}^{N_{F}+1}=A_{ab}\otimes f_{b}^{N_{F}}$
with \vspace{-0.5cm}

\begin{eqnarray*}
A_{ab} & = & \delta_{ab}+\frac{\alpha_{s}}{2\pi}P_{b\rightarrow a}\left[\ln\left(\frac{\mu^{2}}{m_{Q}^{2}}\right)+c_{b\rightarrow a}\right]\end{eqnarray*}
up to $O(\alpha_{s}^{2})$. Here, $P_{b\to a}$ is the DGLAP splitting
kernel and $c_{b\to a}$ is a constant.%
\footnote{The matching conditions are determined entirely by the DGLAP evolution
kernels up to a constant term which must be computed. At NLO, the
constant term is zero such that the PDFs are continuous; at NNLO,
this term is non-zero.%
} The choice $\mu_{M}=m_{N_{F}}$ eliminates the logarithmic terms
thus simplifying the calculation. 

We also observe that shifting the Matching Point from $m_{Q}$ to
$2m_{Q}$ does not suppress the heavy quark PDF as the logarithmic
terms compensate for evolution between $m_{Q}$ and $2m_{Q}$.

\section{Conclusions: }

The computation of heavy quark production has historically been challenging
both theoretically and experimentally. On the theoretical side, the
heavy quark introduces an additional mass scale which complicates
the calculations. On the experimental side, the data for heavy quark
production has typically differed from the theoretical predictions
by a significant factor. Recent theoretical  developments enable us
to incorporate the heavy quark mass into the calculation both dynamically
and kinematically. These calculations have been used to produce matched
PDFs incorporating the full mass dependence. Updated analyses show
improved agreement between data and theory for both HERA and Tevatron
measurements. 

Improved  experimental precision will demand NNLO accuracy from the
theoretical calculations, and this introduces a number of  issues
not present at the NLO order. There is progress underway on both the
PDFs and the hard-scattering calculations, and this should ensure
we are well prepared for the upcoming LHC data. 

\vspace{0.4cm}

\textbf{Acknowledgments:} We would like to thank John Collins, Stefan
Kretzer, Pavel Nadolsky, M.H. Reno, Davison Soper, Robert Thorne,
Wu-Ki Tung, Ji Young Yu, for valuable discussion. F.I.O. acknowledges
the hospitality of CERN, DESY, and the Universit\'e Joseph Fourier,
Grenoble where a portion of this work was performed. This work was
partially supported by the U.S.\,Department of Energy.

\bibliographystyle{plunsrt}
\bibliography{fred}

\end{document}